\documentclass[aip,jap,preprint]{revtex4-1}
\usepackage{graphicx}
\begin{document}
\title{Transport in Graphene Tunnel Junctions}
\author{C. E. Malec and D. Davidovi\'c}
\affiliation{Georgia Institute of Technology, Department of Physics}
\date{\today}

\begin{abstract}
We present a technique to fabricate tunnel junctions between graphene and Al and Cu, with a Si back gate, as well as a simple theory of tunneling between a metal and graphene.  We map the differential conductance of our junctions versus probe and back gate voltage, and observe fluctuations in the conductance that are directly related to the graphene density of states. The conventional strong-suppression of the conductance at the graphene Dirac point can not be clearly demonstrated, but a more robust signature of the Dirac point is found: the inflection in the conductance map caused by the electrostatic gating of graphene by the tunnel probe. We present numerical simulations of our conductance maps, confirming the measurement results.  In addition, Al causes strong n-doping of graphene, Cu causes a moderate p-doping, and in high resistance junctions, phonon resonances are observed, as in STM studies.
\end{abstract}

\maketitle

\section{Introduction}
The electron and hole bands of pristine graphene exhibit linear dispersion, and meet at a point in reciprocal space of zero energy and vanishing density of states (DOS) known as the Dirac point.  This situation leads to charge carriers that behave as massless relativistic particles with a Berry's phase of $\pi$,~\cite{novoselov,zhang} and a large number of theoretical and experimental studies have investigated the consequences of this fact.  The realization of high electronic mobilities~\cite{bolotin,berger,geim} has in turn spurred interest both in fundamental quantum hall studies~\cite{novoselov,zhao,du} and the use of graphene as a next generation electronic material, stable on nm length scales,~\cite{li2,berger} and able to carry extremely high frequency signals~\cite{lin}.

In traditional silicon based electronics, the contacts between semiconductors and other materials play a crucial role. In graphene, however; a prototypical device would be carved out from a single graphene sheet, including the interconnects.~\cite{geim2}, and this is perhaps the reason that the study of contacts between graphene and metals has not received as much attention as the intrinsic effects.
Both for reasons of protection, and the layering necessary to create electronic architectures sufficiently complex for modern integrated circuits, graphene must have other materials stacked on top and possibly beneath it.  Thus the study of heterostructures formed between graphene, and oxides, metals, semiconductors, or even other pieces of graphene is of great technological importance, yet there have been relatively few theoretical~\cite{giovannetti,ran,barraza} and experimental~\cite{huard,venugopal,pi} investigations of such contacts.

In this work, we study Al and Cu junctions separated from graphene by a sufficiently thin oxide to constitute a tunneling junction, and in many devices sufficiently thick to operate as a field effect transistor.
Until now, most low temperature tunneling studies of graphene have been done using scanning tunneling microscopy (STM).~\cite{deshpande2,zhang3,zhang4} Those studies provided a wealth of information regarding the electronic states in graphene,~\cite{martin,li,miller} the Landau levels in a magnetic field,~\cite{li,miller} and inelastic tunnel processes.~\cite{zhang} The few implementations of solid state junctions so far have almost exclusively focused on spin-injection studies.~\cite{han,tombros}

\section{Fabrication and Experiment}
\begin{figure}
\includegraphics[width=0.75\textwidth]{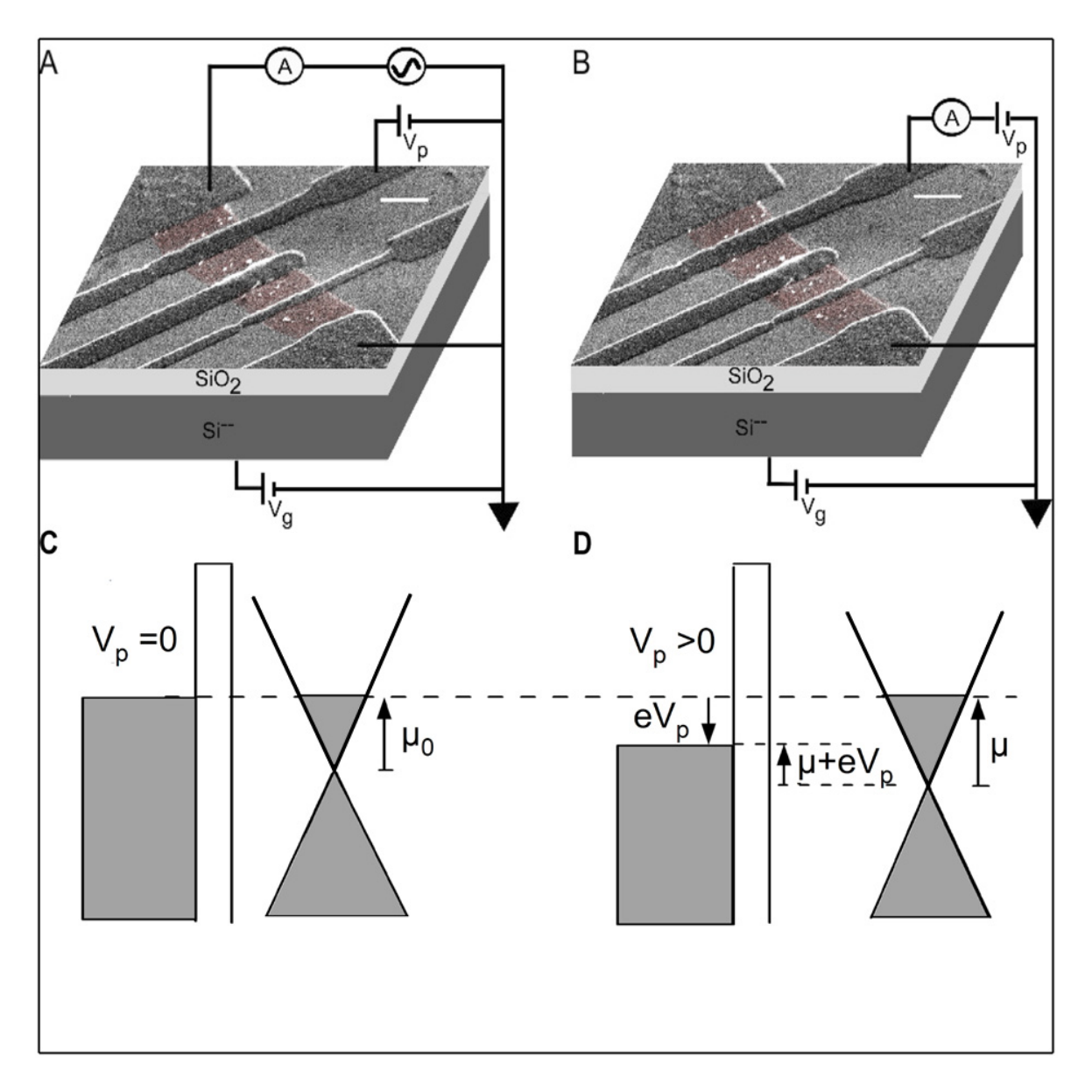}
\caption{Colorized SEM image of device S1 (after measurement) with schematic of A: an FET experiment and B: a tunneling experiment.  For low resistance samples, a three-probe measurement was used to remove the effect of the contact resistance of the grounding electrode. Scale bar is 1$\mu$m. C: Schematic of Fermi levels at zero and D: non-zero probe voltage\label{fig1}}
\end{figure}

We fabricate junctions by visually selecting samples of graphene exfoliated onto an oxidized silicon wafer that is used as a back gate.  We confirm that they are a single graphene layer using raman spectroscopy~\cite{ferrari}. We use electron-beam lithography to define a device pattern in a bilayer of PMMA/MMA resist, we then thermally evaporate 35 nm of metal, and lift-off in acetone.  An image of a finished device is pictured in   Fig.~\ref{fig1}.  The large area ($>$5$\mu$m$^2$) leads have contact resistances to graphene of order k$\Omega$; they will be referred to as the ohmic leads. In Al, the leads $\sim$.5$\mu$m wide or less rapidly increase their contact resistance to graphene after removal from the vacuum chamber. In Cu, leads are made narrower (100-200 nm), and placed in dilute HNO$_3$ (12\% in water) for several seconds to initiate the oxidation.  The devices are rinsed in DI water, IPA, and blown dry.  At this point they begin to age in air, though more slowly than the Al junctions. The resistance of the junction reaches the $M\Omega$-range after  aging for several days. It is essential not to heat the Cu samples above room temperature, during the aging or before measurement (heating can promote rapid oxidation in air, and vacuum heating above $\sim$100 C will reduce the Cu-oxide). In this letter we will demonstrate that these narrower, oxidized leads are tunneling probes of graphene.

We suppose that due to these metals' poor bonding with graphene,~\cite{giovannetti} diffusion of oxygen and water through the interface is responsible for the oxidation of the junction. This method has been used for fabricating tunnel junctions in the past.  In his original tunneling experiments, Giaever ~\cite{fisher} oxidized Al in normal air to create tunnel junctions of $\sim$50\AA~ thickness, and the procedure worked even when a semiconducting film was deposited on top of the metal electrode before removal from the vacuum chamber.~\cite{giaever} It should be noted that a similar method was used to study screening in multilayer graphene devices,~\cite{miyazaki} and that similar reasoning was applied to the poorly bonding metal gold to make tunnel junctions to nanotubes.~\cite{bachtold}

After allowing the junction to age in air to a high resistance, we cool the device to 77K to arrest the aging process.  At low temperatures, the devices can sit for many days without noticeable change to their resistance, whereas at room temperature they may increase several orders of magnitude within several hours.  The junction resistance changes by about 10\% upon cooling room room temperature to 77K, and by less than an order of magnitude upon cooling to 4.2K.  All transport measurements in this report are performed at 4.2K.

We perform two complementary graphene studies. First, we use the probe as a local gate in sufficiently high resistance junctions.  The experimental setup is shown in Fig.~\ref{fig1}-A. By measuring the change in resistance between the two ohmic leads, we obtain the average doping in graphene under the probe.

In the second experimental setup, the current between the probe and the graphene was measured versus probe voltage, while the back gate voltage is swept slowly. The measurement geometry is shown in Fig.~\ref{fig1}-B. We then numerically calculate a derivative from the IV curve, and create a 2D map of differential conductance versus probe ($V_p$) and back gate voltage ($V_g$).  These 2D maps will be referred to as conductance maps.

We have investigated many samples. We focus on our highest quality samples which include an Al junction tested at several points in the aging process (Al-1A, Al-1B, and Al-1C), a second Al sample (Al-2), and two Cu junctions (Cu-1 and Cu-2).  A few other samples have been studied in less detail, but generally confirm the effects presented here. The sample displayed in Fig.~\ref{fig1} is Al-1.

\section{Theory of Tunneling in Graphene}\label{theory}

We now develop a quantitative theory which we can use to analyze our data.  Since the graphene DOS is much smaller than in metals, it can not be assumed that the probe electrode used for tunneling does not itself gate the graphene.  This situation merits a rethinking of some standard analysis.  We use as a starting point the expression for a tunneling current into a two dimensional material.  When tunneling into very thin films, the electron tunnels from a bulk metal, into a state which is confined in the direction perpendicular to the plane of the electrode.  Thus there are electron in a box states in the perpendicular direction.  Since graphene is nearly two dimensional, these states are exceptionally widely spaced (many volts), and in practice only the very first one can actually be accessed.  Therefore, taking $E_{nx}$ as the energy of the n$^{th}$ electron in a box state, $D(E,E-E_{nx})$ as the transmission coefficient at an energy $E$ and a perpendicular energy $E_{nx}$, and $g_n(E-E_{nx})$ as the density of states (DOS) of the n$^{th}$ band, we have for the tunneling current at zero temperature~\cite{wolf}

\begin{equation}
\label{current}
J_n=\frac{e}{h}\int_{\mu}^{\mu+eV_p}g_n(E-E_{nx})D(E,E-E_{nx})dE
\end{equation}

We assume that only n$=1$ is important, and that $D(E,E-E_1)=D$.  We justify the assumption of a constant transmission coefficient of the barrier in the section on inelastic tunneling, in particular, that the barrier height is much larger than the bias voltage.  We, however; do not assume that $\mu$ stays constant while the probe voltage is changed. Then,

\begin{equation}
\label{conductance}
G=\frac{dJ}{dV_p}=\frac{De^2}{h}\left[g(\mu+eV_p)\left(1+\frac{1}{e}\cdot\frac{d\mu}{dV_p}\right)-\frac{g(\mu)}{e}\cdot\frac{d\mu}{dV_p}\right]
\end{equation}

where $g(E)=g_{1}(E-E_1)$.

Normally, when looking at the tunneling conductance, the Fermi level is taken as constant, but  in our junctions, $\mu$ is a function of both $V_p$ and $V_g$, and we use a capacitor model to express the relationship. The graphene chemical potential $\mu$ is measured from the Dirac point, as sketched in  Fig.~\ref{fig1} C and D. Graphene is grounded via the Ohmic contact shown in  Fig.~\ref{fig1} A or B. In equilibrium ($V_p=V_g=0$), the graphene electrochemical potential, $\mu+e\phi$, is equal to that of ground, which we set to zero, $\phi$ is the electrostatic potential, and $e=-1.602\cdot 10^{-19}$C. Since the gate is well insulated from graphene, the graphene remains in equilibrium with ground at finite $V_g$.
 At finite $V_p$, the current flows between the probe and the graphene, resulting in a potential drop across the graphene and the Ohmic contact. We can neglect this potential drop,  because the resistance between the tunnel probe and graphene (typically $>M\Omega$) is much larger than the sum of the graphene resistance and the Ohmic contact resistance (typically $< 10k\Omega$), so the condition $\mu+e\phi=0$ remains satisfied in graphene. 
 
 At finite probe and gate voltages, electrons will be attracted or repelled from graphene, because of the capacitive coupling to the probe and the gate. The change in the carrier density changes the graphene chemical potential, while the electrostatic potential adjusts to maintain equilibrium $\mu+e\phi=0$. In our capacitor model,

\begin{equation}
\label{capacitance}
C_p(V_p-\phi)+C_g(V_g-\phi)=-\sigma_0-e\int_{\mu_0}^{\mu}g(E)dE
\end{equation}

where $C_p$ and $C_g$ are the capacitances between the probe and graphene, and the gate and graphene, respectively, per unit area,
and $\sigma_0$ and $\mu_0$ are the charge density  and the chemical potential at $V_g=V_p=0$, respectively. It follows that $\sigma_0=-(C_p+C_g)\mu_0/e$ for the initial doping of graphene. C$_g$ is measured to be 124$\mu$F/m$^2$ on a test sample. 

In ideal graphene, $g(E)=g_0(E)=D_0|E|$, where $D_0=2/\pi(\hbar v _F)^2=1.47\cdot 10^{18}/(eV\,m)^2$ is the DOS per unit energy and unit area, assuming $v_F=1.0\cdot 10^6$m/s. As a function of $V_p$ and $V_g$, the resistance of graphene has a maximum when the Fermi level is at the Dirac point, $\mu=0$. Substituting $\mu=0$ and $g(E)=g_0(E)$
in  Eq.~\ref{capacitance}, and solving for $\mu_0$, we obtain

\begin{equation}
\mu_0=-sgn(C_{p}V_{p,D}+C_gV_{g,D})\left[-\frac{C_p+C_g}{D_0e^2}+\sqrt{\left(\frac{C_p+C_g}{D_0e^2}\right)^2+ \frac{2|C_pV_{p,D}+C_gV_{p,D}|}{|e|D_0}}\right],
\label{FET}
\end{equation}

where $V_{p,D}$ and $V_{g,D}$ are the probe and the gate voltages, respectively, at the resistance maximum.  We will use this relation to obtain the average doping in graphene under the probe and outside the probe.

\begin{figure}
\includegraphics[width=0.75\textwidth]{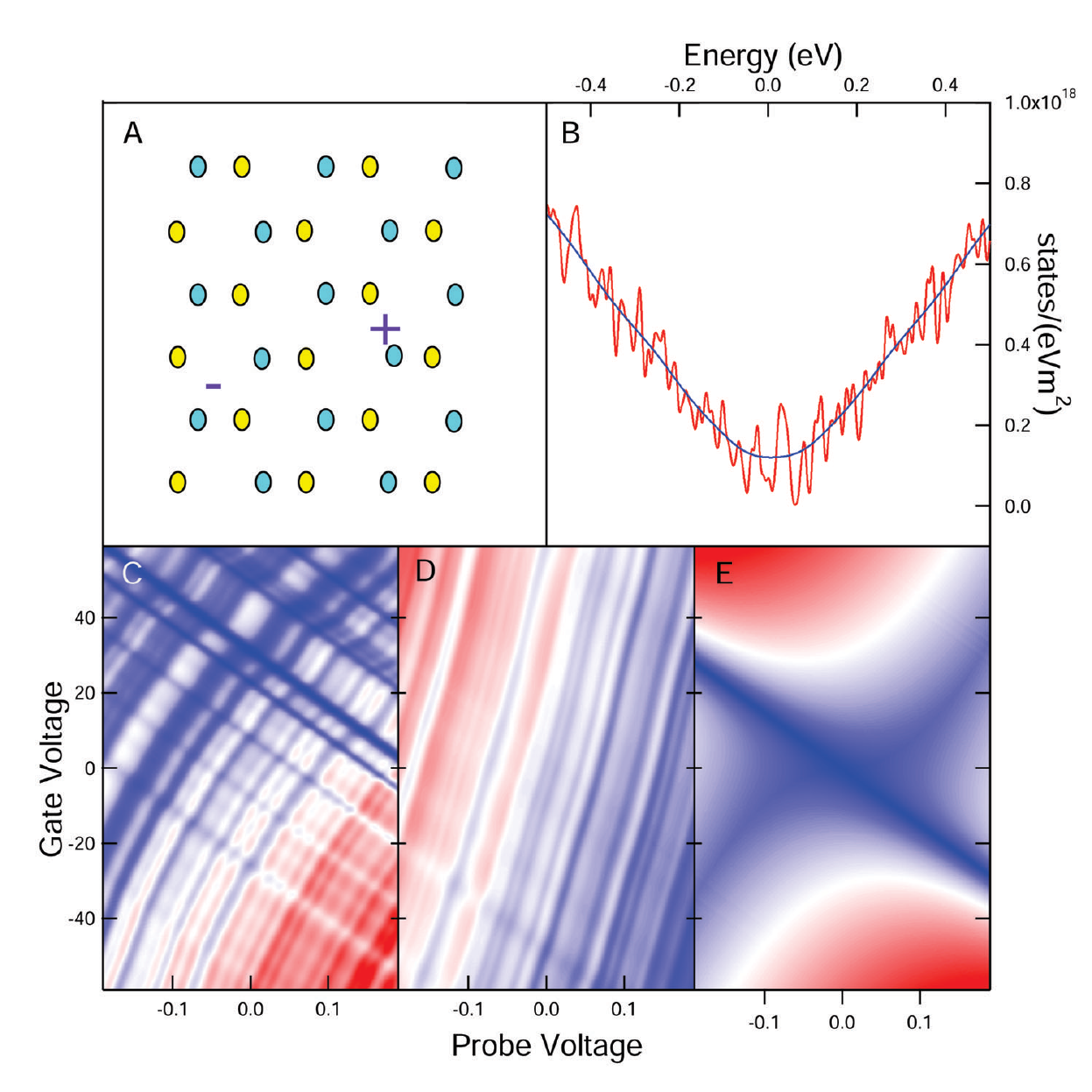}
\caption{A: The graphene lattice with randomly placed impurities B: calculated DOS with 5 meV smoothing function (red), and 50 meV (blue) C: Simulated conductance map with .150 eV p-doping D: A conductance map of the same DOS with .4 eV n-doping. E: Simulation of conductance map for the ideal graphene density of states.  There are two suppressions in the conductance as the probe energy (parabolic line), and fermi energy (negative sloping line) pass through the Dirac point.\label{fig2}}
\end{figure}

Though a great deal can be drawn from these analytic equations, and they can be explicitly solved for the ideal graphene DOS, $2|E|/\pi(\hbar v_F)^2$, they can become rather complicated if one tries to insert a non-ideal value for the DOS, and thus we have set out to calculate conductance maps numerically.  We use a tight binding hamiltonian with periodic boundary conditions and a potential of randomly placed charged impurities, sketched in  Fig.~\ref{fig2}-A. The Hamiltonian is then written as~\cite{pereira2}
\begin{equation}
H=\sum_i^N E_ic_i^{\dag}c_i + \sum_{i,j}^{n.n.}t_{ij}c_j^{\dag}c_i + \frac{Ze^2}{4\pi\epsilon}\sum_{i}^N\sum_k^{N_{imp}}\frac{c^{\dag}_ic_i}{r_{ik}}
\end{equation}
where $t_{ij}=t=2.25$eV is the hopping energy, $E_i=E=0$ is the site energy, and $N_{imp}$ corresponds to an impurity density of $\sim 3\cdot10^{11} cm^{-2}$.  The hopping energy is chosen so that the final result for the Fermi velocity agrees with $1.0\cdot 10^6m/s$. We use an even number of oppositely charged impurities so that there is no net doping, and we may enforce a doping of our choosing ($\sigma_0$ in  Eq.~\ref{capacitance}) when constructing a conductance map.  We simulate an area of graphene corresponding to roughly 11,000 unit cells, or 600 nm$^2$.  The eigenvalues $E(\alpha)$ are derived from the matrix, the DOS is obtained as $g(E)=\sum_{\alpha}\delta(E-E(\alpha))$, and a $5meV$ broadening was introduced to the eigenvalues to create a continuous DOS.

 Fig.~\ref{fig2}-B displays a typical DOS, indicating the fluctuations with energy. Also shown is the same spectrum with 50 meV broadening, showing the correct functional dependence.  In the vicinity of the Dirac point, the fluctuation amplitude is comparable to the average DOS.
The fluctuations will prevent clear observation of the tunnel-conductance suppression near the Dirac point. It is important to understand that far away from the Dirac point, the fluctuations are weak and the density of states is close to linear.  Realistically, the coulomb interactions in graphene are screened~\cite{hu}, but the unscreened coulomb potential used in ref~\cite{pereira2} creates fluctuations in the DOS.  To explain our conductance maps, only the fluctuations are necessary, and their physical origin is a topic for more detailed investigation.  In fact it would have been possible to produce the same qualitative conductance diagrams just by adding noise to an ideal graphene DOS.

 To obtain a conductance map, we solve Eq.~\ref{capacitance} numerically to obtain $\mu$ and $d\mu/dV_p$ as a function of $V_p$ and $V_g$, and calculate the conductance with  Eq.~\ref{conductance}.   A conductance map constructed at $\mu_0=-.15eV$ and $\mu_0=+.4eV$ doping is displayed in  Fig.~\ref{fig2}-C and D repsectively.  Clearly, two sets of parallel lines, one with positive slope and one with negative, are present in the maps.

The positive and negative sloping lines in the conductance map
arise when $\mu+eV_p=const$ and  $\mu=const$, respectively. For example, if $V_p$ and $V_g$ are varied so that $\mu+eV_p=const$, the upper bound of the integral in  Eq.~\ref{current} is constant, so the contribution to the differential conductance from the upper bound is constant. A similar condition holds for the lower bound, if $V_p$ and $V_g$ are varied so that $\mu$ is constant.

To see how this translates into the lines we see in the conductance maps, we take the differential of  Eq.~\ref{capacitance}, and finding how the back gate electrode must change to enforce either $\mu + eV_p=constant$ or $\mu=constant$
\begin{eqnarray}
\label{constantenergy}
\frac{dV_g}{dV_p}\biggl.\biggr |_{\mu+eV_p}=1+\frac{e^2g(\mu)}{C_g}=1+\frac{C_Q}{C_g}\\
\label{constantcharge}
\frac{dV_g}{dV_p}\biggl.\biggr |_{\mu}=-\frac{C_p}{C_g}
\end{eqnarray}
where $C_Q$ is the quantum capacitance as defined in~\cite{luryi} replacing the graphene DOS for that of a 2DEG.
We can thus see that the slope of lines of constant energy $\mu+eV_p$ always have positive slope, and are related to the graphene DOS at the Fermi level. The curvature of the positive sloped lines $\mu+eV_p=const$ is
\begin{equation}
\label{constantenergy1}
\frac{d^2V_g}{dV_p^2}\biggl.\biggr |_{\mu+eV_p}=-\frac{e^3}{C_g}\frac{dg(\mu)}{d\mu}\\
 \end{equation}

Fig.~\ref{fig2}-E shows a conductance map calculated for an ideal, undoped graphene DOS, to clearly illustrate how Eq. 6-8 are interpreted. In ideal graphene, there are no DOS fluctuations and there is only one sharp feature in the DOS, the Dirac point. There are two distinct signatures of the Dirac point in the conductance map. One is the conventional conductance suppression at energy $eV_p+\mu=0$, which leads to a positive sloping line in the conductance map. The curvature of the positive sloping line is $sgn(\mu)|e|^3D_0/C_g$. The curvature magnitude is constant except at the inflection point. The curvature is positive or negative, for $\mu >0$ (n-doped) and $\mu <0$ (p-doped), respectively. The second signature of the Dirac point is a conductance minimum along the line of $\mu = 0$. As $V_g$ is varied, the second minimum shifts along the line $C_gV_g+C_pV_p =0$. Thus, for a given gate voltage, the probe voltages corresponding to the two Dirac-point conductance minima have opposite signs and different magnitudes. The two lines intersect at $V_p = 0$.  Recently, an STM experiment on exfoliated, backgated graphene has revealed a conductance map resembling our ideal case ~\cite{jung}, though in making comparisons it should be noted that STM typically uses a bias convention reversed from that presented here.

In finite size disordered graphene, the DOS fluctuations with energy  lead to multiple minima and maxima in conductance near the Dirac point. A set of parallel positive sloping lines in the conductance map arise. Along those lines, $eV_p+\mu$ is constant, equal to the energy of a feature in the graphene DOS. The curvature of the positive sloping lines changes from negative to positive in the vicinity of $\mu=0$.  Among the various positive sloping lines, the inflection shifts following the negative sloped line $C_gV_g+C_pV_p =const$. The negative curvature in the positive sloping lines in Fig.~\ref{fig2}-C indicates p-doping, while the weak positive curvature in the positive sloping line in Fig.~\ref{fig2}-D indicates n-doping. We will show that Fig.~\ref{fig2}-C and Fig.~\ref{fig2}-D agree with the conductance maps of Cu and Al junctions respectively.  The strong suppression in the conductance of ideal graphene is difficult to observe due to the DOS fluctuations near the Dirac point. However, the inflection (flattening) remains visible, since the fluctuations move in parallel despite disorder. Thus the inflection is a more robust signature of the Dirac point in the tunneling data.\footnote{We have also obtained the local DOS at site $i$,  $g_i(E)=\sum <\alpha|c_i^+c_i|\alpha>\delta(E-E(\alpha))$, found the corresponding conductance map (which assumes tunneling into the site $i$ only), and find qualitatively similar behavior as in  Fig.~\ref{fig2} C and D. The local DOS would be appropriate for a pinhole, while the global DOS $g(E)$ would be for a uniform tunnel junction.}

\section{Experimental Results and Discussion}
\subsection{FET geometry}
\begin{figure}
\includegraphics[width=0.75\textwidth]{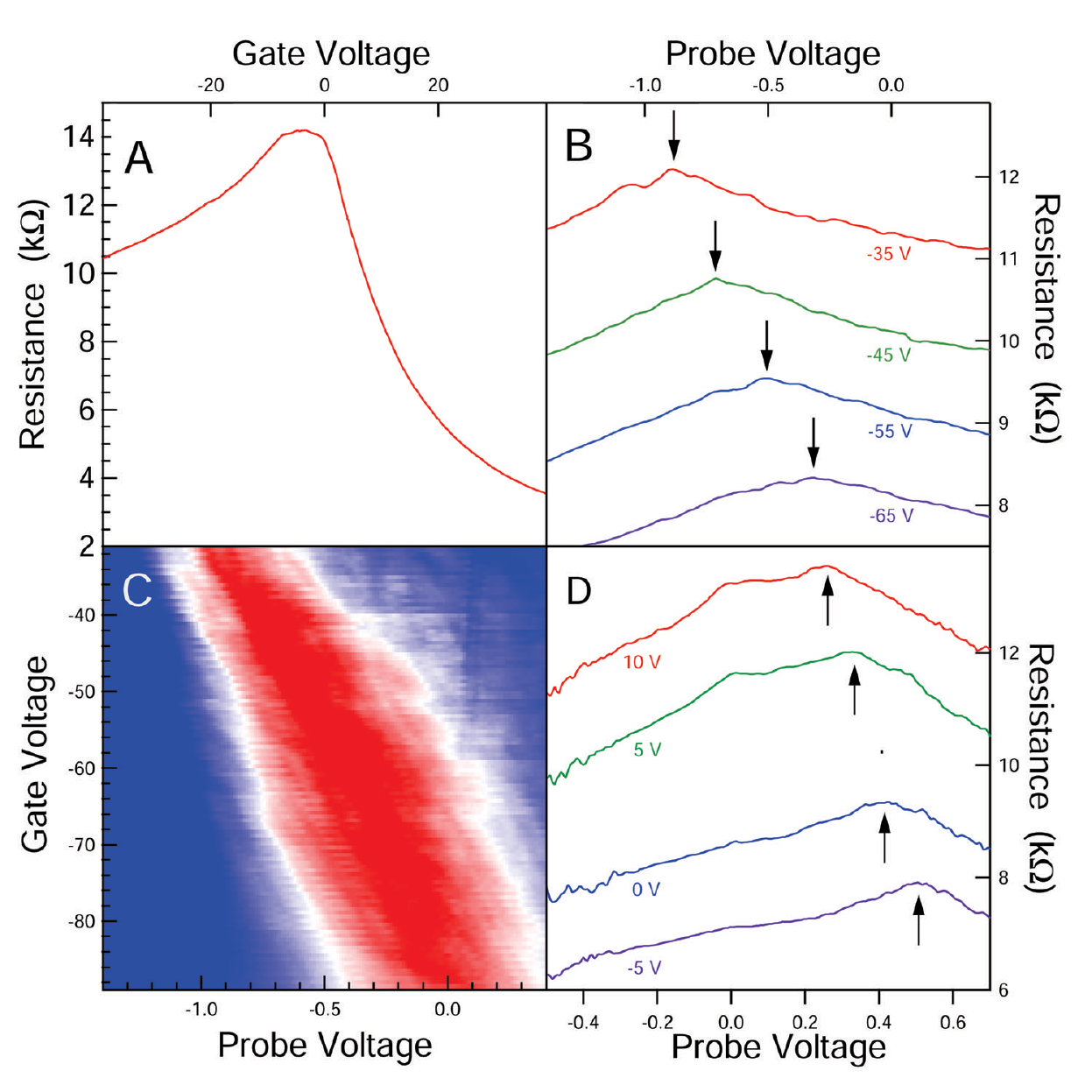}
\caption{A: The back gate voltage response of bulk graphene in Al-1B at $V_p=0$, the Dirac point is near zero, indicating weak doping.  B: Several scans of the probe voltage response at different back gate voltages for Al-1B, the Dirac point can be seen to shift towards zero probe voltage with large negative back gate voltage. C: 2D colorscale of 2 probe resistance with probe and back gate voltages for Al-1B.  The slope of the Dirac point beneath the probe is measured to be 72, and the bulk Dirac peak has been subtracted for clarity. D: Several scans of the probe voltage response at different back gate voltages for Cu-1.\label{fig3}}
\end{figure}

We will now discuss the results from the first setup, sketched in Fig.~\ref{fig1}-A.  This experiment gives us a global view of the capacitance of the probe electrode, and the doping level of the graphene under the probe.  The resistance between the two Ohmic leads, that is the two-probe resistance, was measured using a lock-in technique, with an excitation current of 1$\mu$A.
The two-probe resistance versus back gate voltage ($V_g$), at zero probe voltage, ($V_p=0$), is shown in Fig.~\ref{fig3}-A for sample Al-1B.   The two-probe resistance has a maximum at $V_g=-.4$ volt, indicating a Dirac point in bulk graphene. Since the maximum location is very close to zero volts, this indicates that the doping in graphene is very weak. The doping in this sample is weak by random chance; among different samples, the location of the maximum varies by up to $20$ volts.

Next, the two-probe resistance is measured while sweeping the probe voltage at fixed back gate voltage, as shown in  Fig.~\ref{fig3}-B for sample  Al-1B.  Again, the two-probe resistance exhibits a maximum in resistance, but the location of the peak is very different from that measured in bulk graphene.  The effect of the probe and back gate voltages can be separated by mapping resistance in colorscale versus probe voltage and back gate voltage as seen in Fig.~\ref{fig3}-C. The peak beneath the probe is much smaller due to the smaller area beneath the probe as compared to the area not underneath the probe, so the bulk Dirac peak has been subtracted from  Fig.~\ref{fig3}-C.  There is no visible effect on the bulk peak by the probe voltage.  Similar patterns have been seen in other double gated samples.~\cite{huard2,kim} For Cu samples, the two-probe resistance also exhibits a maximum in resistance versus $V_p$, as shown in  Fig.~\ref{fig3}-D for sample Cu-1, but the maximum is located at positive probe voltage.

The slope of the resistance maximum in  Fig.~\ref{fig3} B,C, and D indicates the capacitance ratio of the probe with the graphene and back gate with graphene, and the intercept of the peak at zero back gate voltage is the Dirac point underneath the junction. The capacitance ratio for the tunnel probe and the back gate is obtained by locating the maximum in the curves shown in Fig.~\ref{fig3}-B and D and finding the best linear fit of the maximum location versus $V_g$. This capacitance ratio gives us a global measure of the junction thickness, and is 71.9 in Al-1B, 32.2 in Al-1C, 73.2 in Al-2, and 85 in Cu-1. We will discuss the capacitance ratio again, after discussing the tunnel spectroscopy results that lead to a higher capacitance ratio. 

In the case of bulk graphene (outside the probe), $C_p\ll C_g$ and we obtain $\mu_0=+0.020eV$, indicating weak doping.   Beneath the probe, for sample Al-1B we obtain $\mu_0=+.25eV$, sample Al-1C we obtain $+.22eV$, sample Al-2 $+.29eV$, and Cu-1 is $-.15eV$.  We would like to stress that all Al junctions tested are n-doped, all Cu junctions are p-doped, and that this doping differs from that in the bulk graphene.  This claim will be further verified by the tunneling spectra.  Note also that this is a global measure of the doping under the probe. That is, $\mu_0$ represents the average doping level over the entire area under the probe.

\subsection{Tunneling geometry}
\begin{figure}
\includegraphics[width=0.75\textwidth]{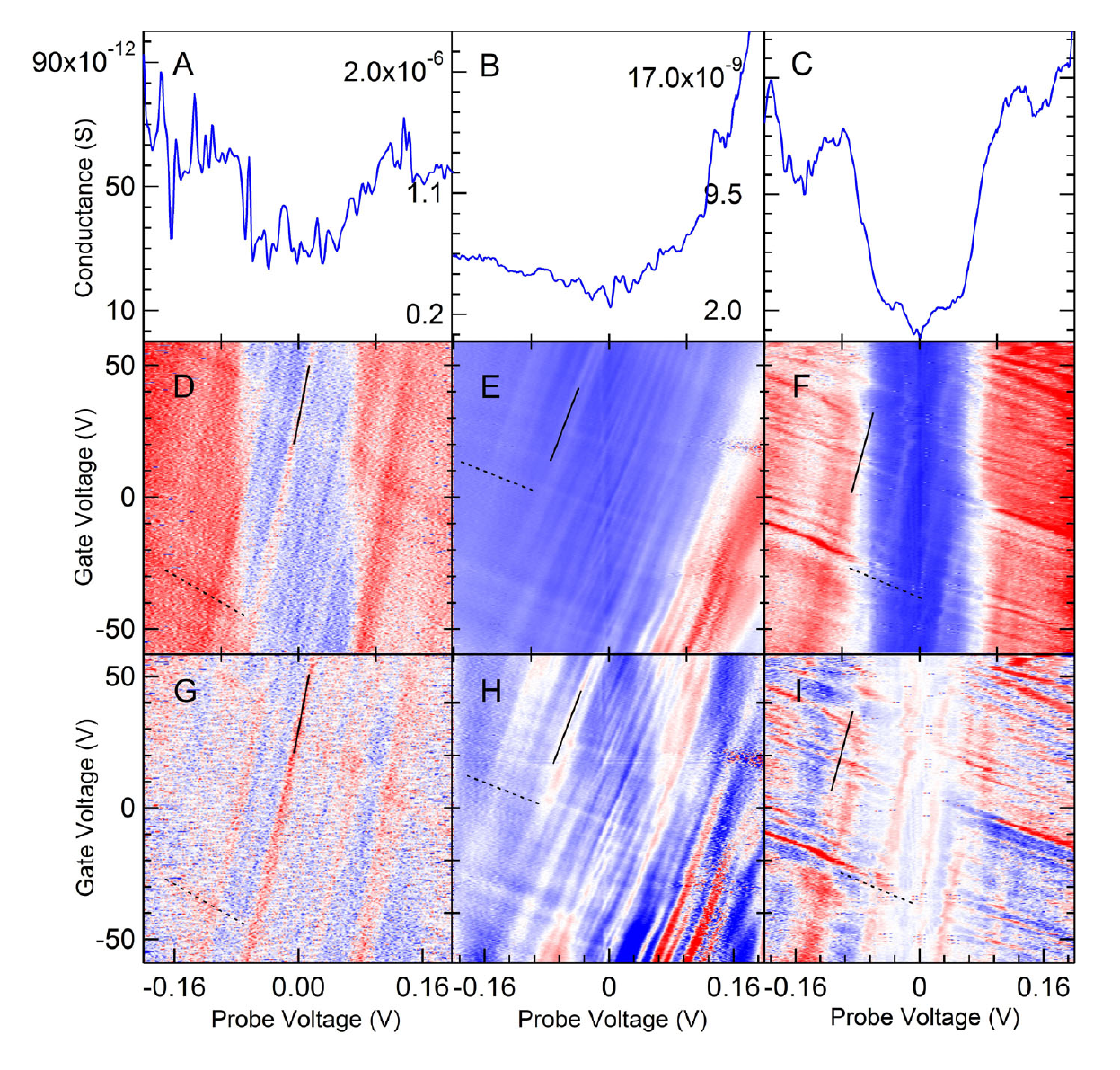}
\caption{A-C: dI/dV vs Probe Bias at zero back gate voltage for three different junction resistances in Al-2, Al-1B, and Al-1C.  D-F: 2D conductance maps of A-C vs probe voltage and back gate voltage.  Two sets of parallel lines are visible. G-I: D-F after subtracting the background conductance, enhancing the two sets of parallel lines.\label{fig4}}
\end{figure}

\begin{figure}
\includegraphics[width=0.75\textwidth]{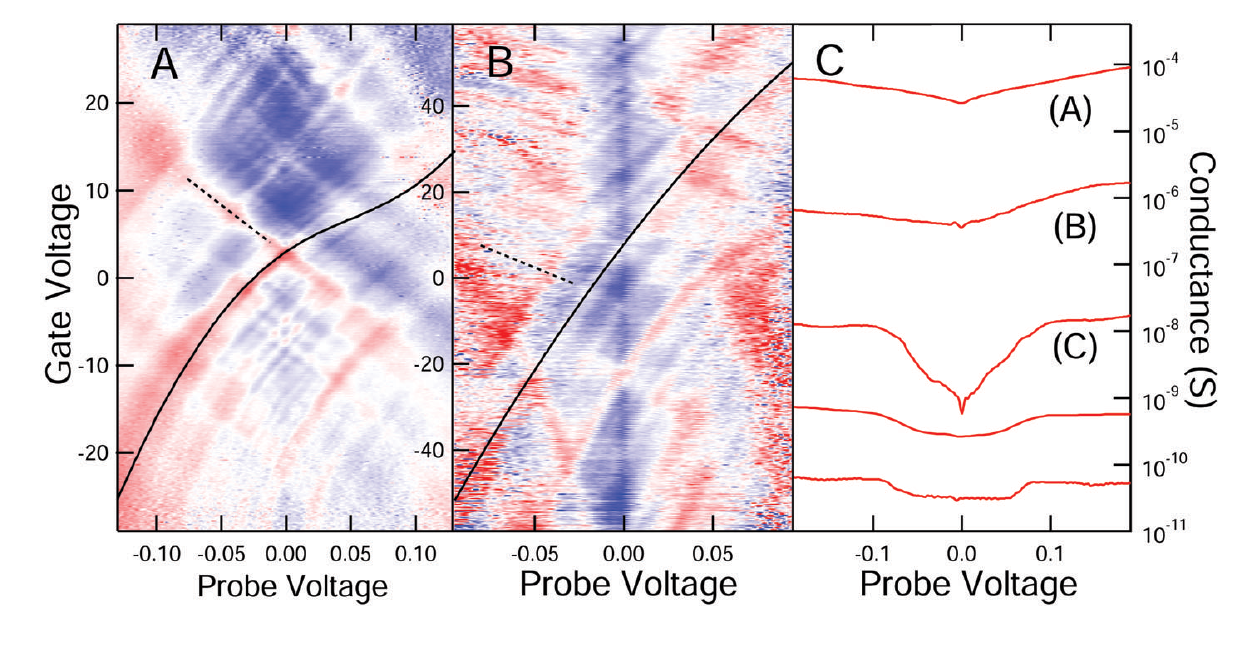}
\caption{A and B: Conductance maps for Cu-1 and Cu-2, respectively.  The superimposed lines are explained in the text. C: Conductance vs probe bias averaged over many back gate voltages, five different Al samples are displayed here, the top three being from Al-1A-C.  The phonon resonance can be seen to emerge at high junction resistances.  \label{fig5}}
\end{figure}

Now we discuss the results from the set-up sketched in  Fig.~\ref{fig1}-B.   Fig.~\ref{fig4}-A, B, and C display the  conductance $dI/dV_p$ versus bias voltage $V_p$ at 4.2K, for sample Al-2, Al-1B, and Al-1C, respectively, at $V_g=0$.  The probe voltage range is smaller than the ranges in  Fig.~\ref{fig3}-B, C, and D, because the current becomes noisy at large bias,  $|V_p|> 0.2V$ . Similarly, the current becomes noisy if the back gate voltage is large, $|V_g|> 60V$.    Fig.~\ref{fig4}-D, E, and F display the conductance maps, defined as $G(V_p,V_g)=dI/dV_p$ as a function of $V_p$ and $V_g$.  As noted above, the figures correspond to the samples in Fig.~\ref{fig4}-A, B, and C, respectively.  Fig.~\ref{fig5}-A and B display the conductance maps for Cu-1 and Cu-2.

The conductance in Fig.~\ref{fig4} exhibits fluctuations, which are reproducible with bias voltage.    Fig.~\ref{fig4}-G through I display the conductance with a low frequency background subtracted, this subtraction greatly enhances the features under discussion.  Fig.~\ref{fig5}-C displays the conductance averaged over all back gate voltage, which demonstrates the emergence of inelastic threshold in large resistance samples.

The central observation of this letter can be seen in the positive and negative sloping lines of Fig.~\ref{fig4}-D through I, and  Fig.~\ref{fig5}-A and B. The positive sloping lines are indicated by full lines, while the negative sloping lines are indicated by dashed lines.  As we explained in Sec.~\ref{theory}, the positive sloping lines are related to the DOS in graphene, while the negatively sloping lines are related to the capacitance ratio between the tunnel and the gate electrode.

The Cu samples in  Fig.~\ref{fig5} display curvature in the positive sloping lines, and in fact all Cu samples tested have displayed curvature in the conductance maps. If the slope of these lines is related to the DOS as we claim, these lines should display an inflection when passing through the Dirac point. We see just this in  Fig.~\ref{fig5}-A, though the lines do not flatten to nearly zero since disorder broadens the DOS.

All Al samples display positive sloping lines that are nearly linear. This would be expected if the doping were high, as in Fig.~\ref{fig2}-D since more voltage is required to shift the Fermi level to change the curvature significantly.  The doping predicted by the FET experiment should allow us to access the Dirac point in the lower portion of some Al junctions, that is, the positive sloping lines should display an inflection in that region. Since no inflection is evident we suppose that the FET experiment is not an accurate predictor of the doping level in the tunnel spectrum.  The discrepancy between the FET results and the tunnel data could be explained by nonuniform doping, which is common in exfoliated graphene.~\cite{zhang3,zhang4} If the tunneling junction were point like (a pinhole), tunneling would depend on local doping, whereas the FET experiment would depend on average doping in graphene under the entire probe.

Thus in  Fig.~\ref{fig4}-G-I and  Fig.~\ref{fig5}-B, doping prevents us from observing the Dirac point within the window of probe and gate voltage which keeps the signal noise acceptably low, though the negative curvature in  Fig.~\ref{fig5}-A and B, and the slight positive curvature in  Fig.~\ref{fig4}-G and H (indicating p and n doping respectively) can clearly be observed.  Despite the lack of a Dirac point in all data, there is still enough information to prove our interpretation.

The lines of constant Fermi level $\mu=const$ are always negative and equal to the capacitance ratio of the two electrodes as discussed in Sec.~\ref{theory}.  This is a local capacitance ratio specific to the small area where the tunneling takes place, and is 160 for Al-1A, 101 for Al-1B, 104 for Al-1C, 133 for Al-2, 120 for Cu-1, and 180 for Cu-2. The fact that this capacitance ratio is much higher than the one stated earlier from the FET measurements indicates that tunneling occurs in a portion of the junction that is thinner than the average junction thickness.  Since the oxidation proceeds inward through the interface, it creates thicker oxide at the edges, and a thinner one near the center.  A thickness measurement of our junctions is not strictly possible using either capacitance value, as we can not know the dielectric constant of the insulating layer.  For the Al-junctions, we can estimate the thickness to dielectric constant ratio from the capacitance of the junction, leading to a typical thickness of 6nm for a dielectric constant of 8.  Since oxide is formed in ambient conditions, with ambient humidity, the dielectric constant could vary from it's ideal bulk value of 8.9. Electron tunneling over such a thick dielectric is not possible, suggesting that any tunneling would be confined to pinholes.

Eq. (6) can also be used to derive the local value of doping in the area that contributes to the tunnel conductance.  By assuming an ideal graphene DOS,
$g_0(\mu_0)=D_0|\mu_0|$, and measuring the positive slopes in conductance map,  we have $+.39eV$ in Al-1B, $+0.76eV$ in Al-1C, and $+.47eV$ in Al-2. 
We can estimate these doping for the Al junctions because at large doping levels the Fermi level doesn't shift much over the conductance map, and the lines remain close to linear. The sign is chosen as positive because of the weak positive curvature.  These doping levels should be compared with those from the FET measurements, $+.25eV$, $+.22eV$, and $+.29eV$, respectively.  Cu junctions, on the other hand, consistently demonstrate positively sloping lines with a curvature described by Eq. (6) and (8).  By adjusting the doping level so that the slope is equal to $1+e^2D_0\mu_0/C_g$ at zero probe and gate voltage, we obtain the correct fit of lines on the conductance map and an estimate of $\mu_0$. The fit lines obtained in this manner are displayed in  Fig.~\ref{fig5}-A and B, and yield doping values of $-.05 eV$ in Cu-1 and $-.25eV$ in Cu-2. The curvature of the best fit lines is independent of the doping level, contains no free parameters, and agrees with the data.

Considering that there is no immediate reason to connect these lines to the graphene DOS, it is remarkable that the positive slopes in the conductance map lead to the Fermi level of graphene close to that in the FET experiment. That is, multiplying the measured positive slope with the back gate capacitance, and dividing it with $D_0$, produces not only the correct order of magnitude, but also the Fermi level that agrees within a factor of two with the Fermi level measured by the FET experiment. The slope of the positively sloping lines {\it is} the DOS of graphene. In addition, the lines of constant energy that display curvature, given by Eq. (6) and (8), agree well with the measurement with only the doping level as an adjustable parameter.

\subsection{Inelastic tunneling}
We now discuss inelastic tunneling.  Fig.~\ref{fig4}-F shows that there are features, at about $\pm 65mV$, that do not shift with back gate voltage. Similar conductance thresholds have been observed in scanning tunnel microscopy of exfoliated graphene \cite{zhang3}. They were attributed to inelastic electron tunneling involving an electron transition in graphene with a phonon emission.  In our samples, the phonon feature is observed only in high resistance samples exceeding $5$M$\Omega$ in tunnel resistance, which restricts the discussion to our Al junctions.  Fig.~\ref{fig5}-C displays the tunnel conductance versus bias voltage in several samples, covering 7 orders of magnitude in conductance. Curves A, B, and C are obtained by averaging the conductance versus $V_p$ over different $V_g$ in Fig.~\ref{fig4}. These curves are obtained in the same sample at various stages of aging. The remaining curves were obtained in different samples. The transition at $\pm 65mV$ is clearly seen in samples with resistance $50G\Omega$, $5G\Omega$, and, $1G\Omega$, while lower resistance samples do not exhibit strong thresholds.

This trend with the tunnel resistance agrees with the explanation from ref~\cite{zhang3,wehling} in terms of inelastic tunneling with an electron-phonon transition.  In an ideal planar graphene tunnel junction, the momentum component of electrons parallel to the plane must be conserved in tunneling~\cite{tersoff}. The Fermi wavevector in Al, $k_F=1.75\cdot 10^{-8}$cm$^{-1}$ is close to the magnitude of the momentum at the K-point in graphene, or $1.70\cdot 10^{-8}$cm$^{-1}$. As a result, one can always find a state on the Al Fermi surface with an in-plane momentum equal to the electron momentum in graphene. In that case, both the in-plane momentum and the energy are conserved in tunneling, which is the reason the elastic tunneling at zero bias voltage is observed in our system.  However, the wavefunctions with large wavevector components parallel to the tunneling interface decay rapidly according to a wave function that behaves like~\cite{tersoff}

\begin{equation}
\psi\sim exp(-(2m_e(V-E_F)/\hbar^2+k_{||}^2)^{\frac{1}{2}}z)
\end{equation}
The $k_{||}$ term increases the effective barrier to tunneling in the case of graphene where $k_{||}$ is the K-point.  For electrons tunneling into the K-point of graphene, $\hbar K^2/2m_e\approx 11eV$.  We believe, therefore, that band bending is not an issue for our bias range of at most 200meV, and is irrelevant in Eq. 4.  However, this does not preclude phonon assisted tunneling.

The energy of the out-of-plane acoustic phonon in graphene with wavevector ${\bf K}$ is approximately $65meV$. So, at a bias voltage of $65mV$ or above, electrons can enter or exit graphene via a higher order, two step process, involving a virtual state at the graphene $\Gamma$-point, which has a wavevector of length zero.~\cite{zhang3}

As the barrier thickness increases by aging, the resistance of the elastic tunnel channel increases much more rapidly than the resistance of the inelastic tunnel channel. In the sample studied in this paper, at the junction resistance of about 5M$\Omega$, the rates of elastic and inelastic tunneling become comparable, and a weak resonance emerges.

It can be seen directly from  Fig.~\ref{fig4}-D,E, and F that the lines are unaffected by inelastic tunneling events, as they can be seen to pass through the inelastic thresholds without changing their slope.  This demonstrates that the inelastic and the elastic tunneling processes are added in parallel, as independent tunneling channels.
\section{Conclusions}
In conclusion, a solid state realization of a tunnel junction into graphene has been demonstrated.  A measurement technique which extracts information of the graphene region under the tunnel junction is introduced. We extract the doping level of the graphene under the junction, both locally at the tunnel spot and globally under the entire junction area, and find doping to be significantly different than in the bulk.  In Al junctions, doping is strong on the n-side, in Cu junctions, doping is small to moderate on the p-side. We observe fluctuations in the tunnel DOS of graphene, which shift with applied back gate voltage in accordance with the electronic DOS in graphene, and an inelastic conduction threshold associated with high resistance junctions. Finally, the electrostatic gating of graphene caused by the tunnel probe needs to be considered, and provides a powerful tool for interpreting tunneling experiments in graphene.
\section{Acknowledgements}
We would like to thank Felipe Birk for assistance in sample fabrication; Salvador Barraza-Lopez, Mei Yin Chou, Phillip First, Walt de Heer, Zhigang Jiang, and Markus Kindermann for useful conversations; and Walt de Heer and Yike Hu for time and assistance with the Raman microscope.  This research is supported by the DOE grant DE-FG02-06ER46281 and David and Lucile Packard Foundation grant 2000-13874.

\newpage

\providecommand{\bysame}{\leavevmode\hbox to3em{\hrulefill}\thinspace}
\providecommand{\MR}{\relax\ifhmode\unskip\space\fi MR }
\providecommand{\MRhref}[2]{%
  \href{http://www.ams.org/mathscinet-getitem?mr=#1}{#2}
}
\providecommand{\href}[2]{#2}

\newpage

\newpage

\newpage

\newpage

\newpage

\end{document}